\documentclass[journal]{IEEEtran}
\usepackage{graphicx}
\usepackage{amsmath, amssymb}
\ifCLASSINFOpdf
\usepackage{float, multirow}
\else
\usepackage{cite}
\fi
\usepackage{xcolor}
\usepackage{cite}
\usepackage{graphicx}
\usepackage{lipsum}
\usepackage{xcolor}
\usepackage{graphics}
\usepackage{hyperref}
\hypersetup{%
  citebordercolor=blue
}
\usepackage{amsmath,amsfonts}
\usepackage{algorithmic}
\usepackage{algorithm}
\usepackage{array}
\usepackage[caption=false,font=normalsize,labelfont=sf,textfont=sf]{subfig}
\usepackage{textcomp}
\usepackage{stfloats}
\usepackage{url}
\usepackage{verbatim}
\usepackage{graphicx}
\usepackage{cite}
\usepackage{dsfont}
\usepackage{xcolor}
\usepackage{balance}
\usepackage{placeins}
\usepackage{subfig}
\begin{document}
\bstctlcite{IEEEexample:BSTcontrol}
\title{Noise Modulation over Wireless Energy Transfer: JEIH-Noise Mod}

\author{
  Erkin Yapici,~\IEEEmembership{Graduate Student Member,~IEEE,} Yusuf Islam Tek,~\IEEEmembership{Graduate Student Member,~IEEE,} \\
  M. Ertug Pihtili,~\IEEEmembership{Graduate Student Member,~IEEE,} Mehmet C. Ilter,~\IEEEmembership{Senior Member,~IEEE,} \\
  and Ertugrul Basar,~\IEEEmembership{Fellow,~IEEE}%
\thanks{Erkin Yapici, Yusuf Islam Tek are with the Communications Research and Innovation Laboratory (CoreLab),  Department of Electrical and Electronics Engineering, Ko\c{c} University, Sariyer 34450, Istanbul, Türkiye.\\
Email: eyapici19@ku.edu.tr, and ytek21@ku.edu.tr}%
\thanks{M. Ertug Pihtili and Mehmet C. Ilter are with the Department of Electrical Engineering, Tampere University, 33720 Tampere, Finland.\\
Email: ertug.pihtili@tuni.fi, and mehmet.ilter@tuni.fi 

E. Basar is with the Department of Electrical Engineering, Tampere University, 33720 Tampere, Finland, on leave from the Department of Electrical and Electronics Engineering, Koc University, 34450 Sariyer, Istanbul, Türkiye. Email: ertugrul.basar@tuni.fi and ebasar@ku.edu.tr }%

\thanks{This work is supported by TUBITAK under Grant Number 124E146.}
}
\maketitle
\begin{abstract}
This letter presents an innovative joint energy harvesting (EH) and communication scheme for future Internet-of-Things (IoT) devices by leveraging the emerging noise modulation (Noise-Mod) technique. The proposed approach embeds information into the mean value of real Gaussian noise samples to enable wireless energy and information harvesting. We propose a mean-based detector to decode the information and derive the analytical bit error probability (BEP) of the proposed scheme under Rician fading channels. We utilize a nonlinear rectenna model to demonstrate the feasibility of the proposed scheme in terms of energy harvesting capability. Simulation results demonstrate that the proposed method outperforms conventional modulation techniques in terms of EH performance across various channel conditions while maintaining reliable communication performance for next-generation IoT networks.
\end{abstract}

\begin{IEEEkeywords}
Noise modulation, energy harvesting, information harvesting, real Gaussian signals, mean-based detection.

\end{IEEEkeywords}

\section{Introduction}

As future Internet-of-Things (IoT) platforms evolve toward sixth-generation (6G), the emphasis on powering IoT devices becomes increasingly crucial, necessitating innovative approaches to deliver power without compromising communication performance \cite{10387322}. Over the past decade, energy harvesting has become a pivotal technology for next-generation communication systems, providing a sustainable solution to meet the increasing power demands of densely deployed low-power IoT devices by enabling simultaneous data reception and powering simple nodes. To support this, simultaneous wireless information and power transfer (SWIPT) is a promising solution for resource-constrained environments through enabling battery-less or battery-assisted devices and paving the way for massive deployments with reduced operational costs \cite{butt2023ambientiotmissinglink, 10586881}. Among the SWIPT techniques, time-splitting (TS) and power-splitting (PS) architectures have been widely studied for their ability to balance energy harvesting and information transmission efficiently. The TS approach divides the transmission interval into distinct time slots, while the PS method splits the received communication signal in the power domain for energy harvesting and information decoding \cite{4595260}. These strategies have demonstrated significant advantages in optimizing the tradeoff between energy efficiency and data reliability in IoT networks \cite{jameel2016optimal}. The recent work \cite{articlezhaoyizhe} investigated energy harvesting modulation for integrated control and power transfer in industrial IoT, underscoring the importance of joint communication and energy harvesting design, which is also targeted by our proposed scheme. 

Noise modulation (NoiseMod) is a novel approach to future wireless communication networks, as it utilizes the statistical properties of Gaussian noise samples, either thermal or artificially generated, as a means of conveying information \cite{10373568,basar2022thercom}. Building on NoiseMod, the authors proposed a noise-domain non-orthogonal multiple access (ND-NOMA) scheme, in which one user's data is modulated using the mean of noise samples, while another user's data is encoded in the variance, effectively leveraging the statistical properties of Gaussian signals \cite{yapici2024noisedomainnonorthogonalmultipleaccess}. The main drawback of the NoiseMod scheme is that it requires a higher sampling rate to achieve data rates comparable to those of conventional schemes. However, its low-power nature makes it suitable for energy-constrained environments, such as Device B types of ambient IoT systems. Moreover, recent studies have demonstrated that Gaussian signals are more suitable for energy harvesting compared to conventional modulation schemes \cite{10586881, Clerckx2016}. In this context, a rectenna, which combines an antenna and a rectifier, plays a key role in explaining the dependency of harvested power on the transmitted waveform. To explore this, in \cite{Kim20220}, the authors proposed a nonlinear rectenna model to investigate the impact of waveform characteristics on the system’s energy harvesting capability. Their findings revealed that real Gaussian (RG) or circularly symmetric complex Gaussian (CSCG) signals can enhance energy harvesting performance. Based on these insights, NoiseMod emerges as a promising modulation scheme for simultaneously delivering power and information to IoT devices.

In this work, we propose a novel NoiseMod-based scheme named joint energy and information harvesting NoiseMod (JEIH-NoiseMod). The proposed scheme operates in a TS manner, where during the communication phase, information is conveyed through the mean of noise samples, enabling information harvesting at the receiver to decode the data embedded in Gaussian samples. In the EH phase, the receiver utilizes the noise samples to power itself. Unlike existing schemes, JEIH-NoiseMod embeds information in the mean of Gaussian noise samples and ensures reliable decoding under Rician fading with free-space path loss considerations, while boosting EH performance to balance the trade-off between harvesting capability and bit error rate (BER) observed in similar systems \cite{Zhou2013} even under imperfect channel state information (CSI). The JEIH-NoiseMod demonstrates strong potential for next-generation IoT networks by transforming undesirable noise into an opportunistic means of power and information transfer. However, noise modulation typically requires a higher sampling rate to maintain comparable data
rates with conventional schemes \cite{10373568}.

\begin{figure*}[t]
    \centering
    \includegraphics[width=1\textwidth]{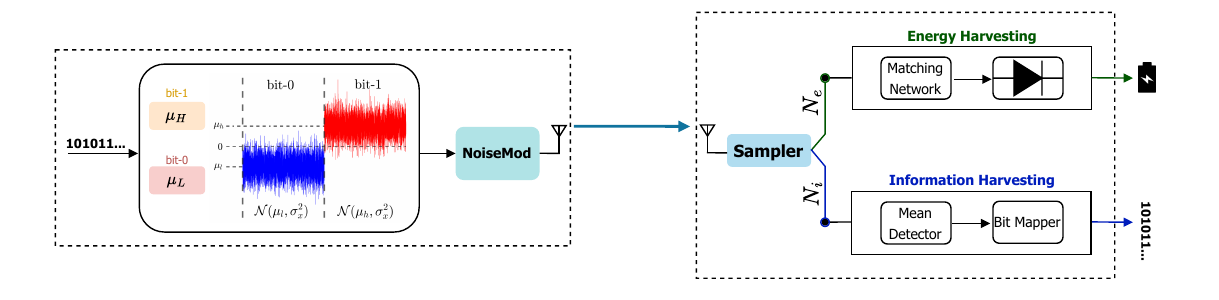}
    \caption{Transceiver scheme of the proposed JEIH-NoiseMod system. Bits are mapped into mean and variance of Gaussian noise in the transmitter and energy harvesting/information harversting are implemented in the receiver, respectively.}
    \label{fig:transceiver}
\end{figure*}
This letter is structured as follows. Section II presents the detailed system model for both information transmission and energy harvesting. Section III provides a theoretical analysis of the BEP expressions. Section IV discusses simulation results for both BER and energy harvesting performance, and Section V concludes the paper.

\textit{Notation:}  \( \mathcal{N}(\mu, \sigma^2) \) denotes the real-valued Gaussian distribution with mean \( \mu \) and variance \( \sigma^2 \), while \( \mathcal{CN}(\mu, \sigma^2) \) represents the circularly symmetric complex Gaussian distribution with mean \( \mu \) and variance \( \sigma^2 / 2 \) per dimension. The expectation operator is denoted by \( \mathbb{E}[\cdot] \), \( I_0(\cdot) \) represents the modified Bessel function of the first kind and order zero, and \( \exp(x) = e^x \) denotes the exponential function with \( e \) as the base of the natural logarithm, and \( |\cdot| \) denotes the amplitude operator for scalar values or the magnitude of a complex number. $\Re(\cdot)$ is the real-part operator.

\section{JEIH-NoiseMod System Model}
In this section, we present the transceiver architecture of the proposed JEIH-NoiseMod scheme that is illustrated in Fig. \ref{fig:transceiver}.  We consider a total number of \( N \) samples per bit duration for joint energy and information harvesting, employing the TS mode as the operation mode \cite{Zhou2013}. In the TS mode, during the first \(N_e = \alpha N \) samples, the information harvester is inactive, and the entire signal power is dedicated to energy harvesting, where $\alpha$ is the portion of time that the energy harvester is active. For the remaining \( N_i = (1-\alpha)N \) samples, the information harvester is active, and all signal power is utilized for information decoding.
In this proposed scheme, the mean value of the RG noise samples is modulated to encode information.  Therefore, this scheme utilizes noise samples with a fixed variance and controllable mean to simultaneously facilitate information transfer and energy harvesting.
\subsection{Information Harvesting}
The transmission of bit-$b$ (bit-0 or bit-1) is realized using \( N_i \) consecutive, randomly generated Gaussian noise samples, and a sample is defined as $x_n \sim \mathcal{N}(\mu, \sigma^2_x)$, , where \( n = 1, \dots, N_i \), $\mu$ and \( \sigma^2_x \) denote the corresponding mean value for bits and the variance of the noise samples (i.e., serves as the overall transmit power factor). $\mu$ is selected as follows:  
\begin{equation}
\mu = 
\begin{cases} 
\mu_l, & \text{if bit-0 is transmitted,} \\ 
\mu_h, & \text{if bit-1 is transmitted,}
\end{cases}
\end{equation}
Herein, $\mu_l = -m$ and $\mu_h = +m$ represent the symmetric mean values corresponding to bit-0 and bit-1, respectively. Under the transmit power constraint, $\Sigma_{n = 1}^{N_i}\mathbb{E}[x_n^2] = \mu^2 + \sigma_x^2 = P_T$, where $P_T$ is normalized to unit power, this symmetric mean selection ensures equal average transmit power for each bit.

Afterward, we express the $n$-th received complex baseband sample $y_n$ over a wireless channel as
\begin{equation}
y_n \;=\; h\,x_n \;+\; w_n,
\end{equation}
where $w_n \sim \mathcal{CN}(0,\,\sigma_w^2)$ corresponds to additive white Gaussian noise (AWGN) sample with variance $\sigma_w^2$. Additionally, for a forthcoming analysis to evaluate the error performance of the system, we define \( \delta = \sigma^2_x / \sigma^2_w \), which represents the ratio of variances between the useful signal and the additive Gaussian noise components. This parameter is analogous to the signal-to-noise ratio (SNR) in conventional digital communication systems. A higher value of \( \delta \) indicates increased robustness of the receiver against the effects of additive disruptive noise, leading to a lower bit error probability (BEP). Moreover, $h=\Bar{h}/{\sqrt{L}}$ is the complex baseband channel coefficient of the transmitter and receiver link. Here, $L$ denotes the attenuation due to the distance between the transmitter and receiver, and it is calculated as
\begin{equation}
L= \left(\frac{4\pi\,d\,f_c}{c} \right)^2,
\end{equation} 
where $d, f_c$, and $c$ correspond to the distance in meters, carrier frequency, and the speed of light, respectively. Furthermore, $\Bar{h}=\Bar{h}_R+j\Bar{h}_I$ is the small-scale channel coefficient for Rician fading with parameter $K$ and $\Bar{h}_R,\Bar{h}_I \sim \mathcal{N}\left(\sqrt{\frac{K}{2(1+K)}},\frac{1}{2(1+K)}\right)$. $\Bar{h}$ is normalized to have unit gain in average, i.e.\ $\mathbb{E}[|\Bar{h}|^2]=1$. Furthermore, we assume block fading, that is $h$ remains constant during each transmission (bit) interval.

At the receiver, the sample mean estimation of the received signals is computed to decode the information bits as
\begin{equation}
   \bar{y} = \frac{1}{N_i}\sum_{n=1}^{N_i} y_n.
   \label{eq:sampleMean}
\end{equation}
The estimated information bit, $\hat{b}$, is detected using the minimum distance rule, which is formulated as
\begin{equation}
    \hat{b} = 
    \begin{cases}
        0, & \text{if} \quad |\Bar{y}-h\mu_l|^2 < |\Bar{y}-h\mu_h|^2 \\
        1, & \text{if} \quad |\Bar{y}-h\mu_h|^2 < |\Bar{y}-h\mu_l|^2. 
    \end{cases}
    \label{eq:Mean_detector}
\end{equation}

\subsection{Energy Harvesting}

In the EH phase, $N_e$ Gaussian noise samples are fed to the rectenna to scavenge energy from received Gaussian signals. The harvested power can then be used directly or stored in a battery. A rectenna, consisting of an antenna and a rectifier, plays a crucial role in the EH process. The rectifier specifically converts the received radio frequency (RF) power into DC power. Herein, we employ a single-diode rectifier model. After passing through a matching network, which ensures impedance matching between the antenna and the rectifier, the single-diode rectifier converts the received Gaussian noise samples into DC power as shown in Fig.~\ref{fig:transceiver}. To account for the practical characteristics of the rectenna based on design preferences, the authors in \cite{Clerckx2016} proposed a nonlinear rectenna model. This model exploits the Taylor series expansion of the diode characteristic function to investigate the effect of different signaling/modulation schemes on the system's energy harvesting capability. This expression is mathematically given by,
\begin{equation}
    z_{\mathrm{DC}} =  k_2 R_{\mathrm{ant}} \mathbb{E}[\lvert y_n \rvert^2] + k_4 R_{\mathrm{ant}}^2 \mathbb{E}[\lvert y_n \rvert^4],
    \label{Eq:z_DC}
\end{equation}
where $k_2$ and $k_4$ are rectifier-dependent parameters, and $R_{\mathrm{ant}}$ represents the antenna impedance. 

Unlike conventional linear rectenna models, where the energy harvesting capability depends only on the received power and not on the utilized waveform, \cite{10586881} highlights the influence of different waveforms on the system’s energy harvesting capability. The term $\mathbb{E}[\lvert y_n \rvert^2]$ represents the linear component and depends exclusively on the received power, while $\mathbb{E}[\lvert y_n \rvert^4]$ accounts for the nonlinear characteristics of the rectenna and varies with the type of transmit signal \cite{Clerckx2016}. Thus, using RG signals instead of CSCG or linear modulated ones is beneficial for energy harvesting, as indicated in \cite{Kim20220}. However, using the mean of RG signals for information harvesting affects the system’s energy harvesting capability. Under the transmit power constraint $\sum_{n = 1}^{N_e}\mathbb{E}[x_n^2] = \mu^2 + \sigma_x^2 = P_T$, where $P_T$ is normalized to unit power, increasing the mean parameter $\mu$ reduces the variance $\sigma_x^2$ accordingly. This reduction diminishes the amplitude fluctuations of the RG signal, thereby reducing the energy harvesting capability, whereas waveforms with amplitude fluctuations are beneficial for enhancing energy harvesting capability \cite{10586881, Clerckx2016}.

\section{Error Analysis}
In this section, we derive the theoretical BEP expressions for the proposed scheme. We assume that the channel is perfectly known at the receiver in this analysis, as well as in computer simulations.

At the receiver, we can further express the sample mean $\Bar{y}$ in \eqref{eq:sampleMean} as
\begin{equation}
\Bar{y}
\;=\;
\frac{1}{N_i}\sum_{n=1}^{N_i} y_n
\;=\;
h\Bar{x}
\;+\;
\Bar{w},
\end{equation}
where $\Bar{x} = \frac{1}{N_i}\sum_{n=1}^{N_i} x_n$ and $\Bar{w} = \frac{1}{N_i}\sum_{n=1}^{N_i} w_n$.
In order to simplify the analysis, we exploit a standard trick of removing the complex phase of the channel. Let us rewrite complex baseband channel coefficient $h$ as $h = r\,e^{j\phi}$, and $\phi$ is the phase of the channel. By removing the phase, we have 
\begin{subequations}
\begin{eqnarray}
    \widetilde{y} \;=\; e^{-j\phi}\Bar{y}, \\
    \widetilde{w} \;=\; e^{-j\phi}\Bar{w}, \\
    r \;=\; e^{-j\phi}h. 
    \label{eq:rotateOut}
\end{eqnarray}
\end{subequations}
Since AWGN is circularly symmetric, $\widetilde{w} \sim \mathcal{CN}(0,\sigma_w^2/N_i)$ has the same distribution as $\overline{w}$. We have the rotated sample mean signal $\widetilde{y}$, which can be expressed as follows
\begin{equation}
    \widetilde{y} = r\Bar{x} + \widetilde{w},
\end{equation}
where $\Bar{x}$ follows $\mathcal{N}(\mu_l,\sigma^2_x/N_i)$. For $b=0$, an error occurs when error event below holds, 
\begin{equation}
    | \widetilde{y} - r\mu_h |^2 < | \widetilde{y} - r\mu_l |^2.
    \label{eq:ErrEvent}
\end{equation}
It can be shown that the error event in \eqref{eq:ErrEvent} can be rewritten as a threshold condition on a real Gaussian variable
\begin{equation}
z \;=\; r\,\left(\overline{x}-(-\,m)\right)\;+\;\Re\{\widetilde{w}\}.
\end{equation}
Let us define an auxiliary variable such that $\omega = \mu_h - \mu_l = 2m$.  Then $z\sim \mathcal{N}\!\bigl(0,\,\tfrac{r^2\sigma^2_x}{N_i} + \tfrac{\sigma_w^2}{2N_i}\bigr)$.
  Conditioned on amplitude~$r$, the instantaneous probability of error is
\begin{equation}
\label{eq:PerrorConditional}
 \Pr(\text{error} \mid b=0, r)
 = \Pr\bigl\{ z > \tfrac{1}{2}r\omega \bigr\}
 = Q\left(
   \frac{\tfrac{1}{2}r\omega}{
      \sqrt{ \tfrac{r^2\sigma^2_x}{N_i} + \tfrac{\sigma_w^2}{2N_i}}
   }
 \right).
\end{equation}
By symmetry, the same holds for $b=1$. Since $\Bar{h}$ has unit second moment, $|\Bar{h}|$ follows a Rician distribution with probability density function (pdf)
\begin{equation}
  f_{|\Bar{h}|}(x) 
  \;=\;
  \frac{x}{\sigma_s^2}\,
  \exp\!\left(
     -\,\frac{x^2 + s^2}{2\,\sigma_s^2}
  \right)\,
  I_{0}\!\left(\frac{x\,s}{\sigma_s^2}\right),\quad x\ge 0,
\end{equation}
where $s^2 + 2\,\sigma_s^2 = 1$ and $K = \frac{s^2}{2\,\sigma_s^2}$.  With path loss $L$, we get $r={x}/{\sqrt{L}}$. Therefore, the scaled Rician pdf can be expressed as
\begin{equation}
\label{eq:ScaledRicianPDF}
f_r(r)
=\;
  \left(\frac{\sqrt{L}\,r}{\sigma_s^2}\right)\,
  \exp\!\left(-\,\frac{L\,r^2 + s^2}{2\,\sigma_s^2}\right)\,
  I_0\!\left(\tfrac{r\,\sqrt{L}\,s}{\sigma_s^2}\right),\quad r\ge 0.
\end{equation}
\begin{figure}[t!]
    \centering
    \includegraphics[width=0.38\textwidth]{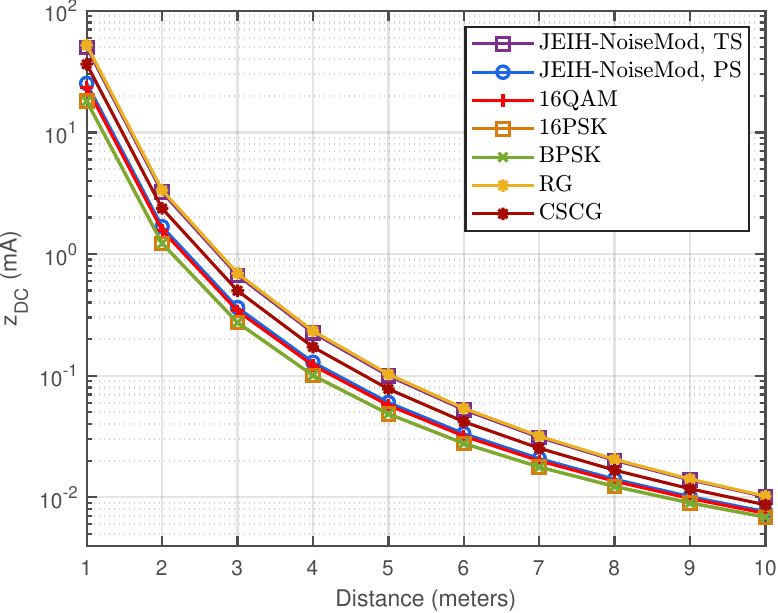}
    \caption{Energy harvesting capability ($z_{DC}$) vs. distance for different modulation schemes.}
    \label{fig:energy_harvesting}
\end{figure}
\begin{figure}[t!]
    \centering
    \includegraphics[width=0.38\textwidth]{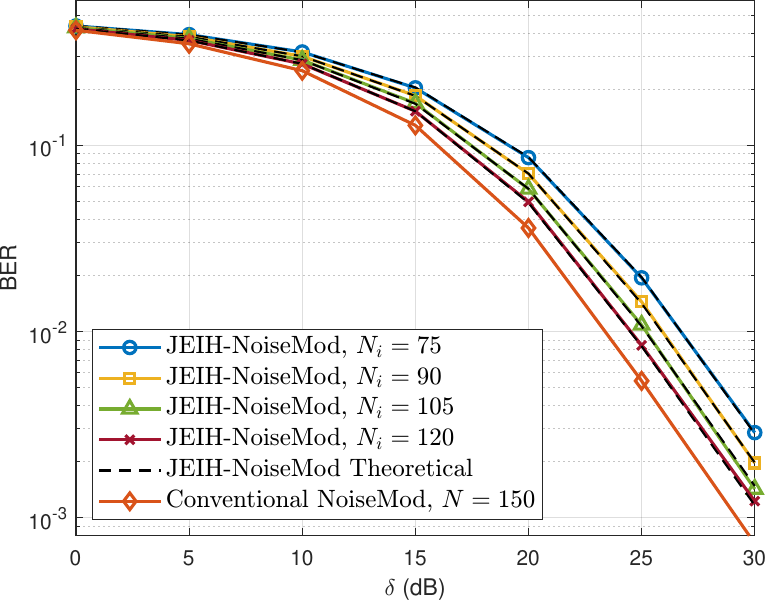}
    \caption{$\delta$ versus theoretical and simulated BER for varying $N_i$ values .}
    \label{fig:BER_Ni}
\end{figure}

Finally, we average $Q$ function in \eqref{eq:PerrorConditional} over the distribution of $r$, and the unconditional BEP is found below
\begin{equation}
\label{eq:PerrorFinal}
  P_{\mathrm{e}}
  \;=\;
  \int_{0}^{\infty}
    Q\left(
      \frac{rm}{
        \sqrt{\frac{1}{2N_i}\left( 2r^2\sigma^2_x + \sigma_w^2 \right)}
      }
    \right)\;
    f_{r}(r)\;\mathrm{d}r.
\end{equation}

We numerically evaluate the integral in \eqref{eq:PerrorFinal} with the adaptive quadrature method \cite{press2007numerical}.
\section{Simulation Results}
In this section, we validate the analytical BER derivations and show the energy harvesting capability based on (6) through computer simulations. The theoretical BEP curves are obtained from \eqref{eq:PerrorFinal}. Table \ref{tab:simulation_parameters} summarizes the simulation parameters, including antenna impedance, rectifier coefficients, carrier frequency, and TS factors.
\begin{figure*}[t]
    \centering
    \begin{minipage}{0.33\textwidth}
        \centering
        \includegraphics[width=\textwidth]{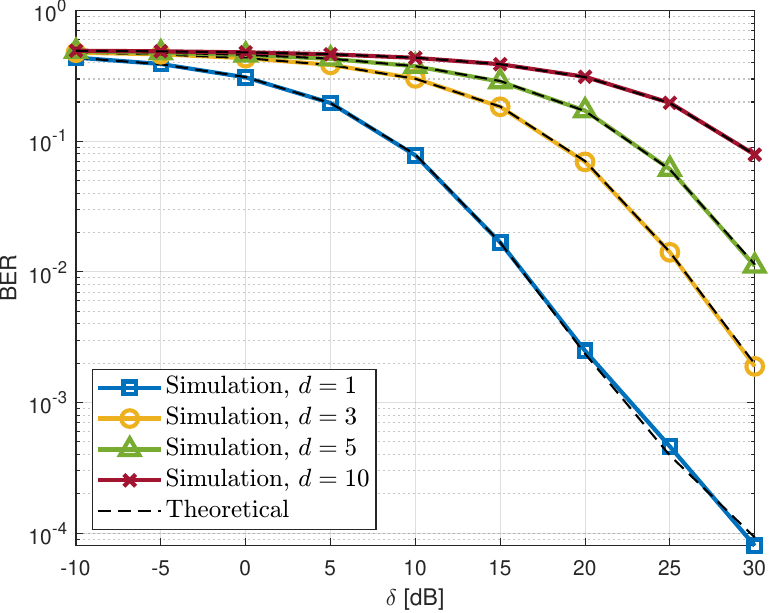}
        \text{(a)}
        \label{fig:ber_d}
    \end{minipage}%
    \hfill
    \begin{minipage}{0.33\textwidth}
        \centering
        \includegraphics[width=\textwidth]{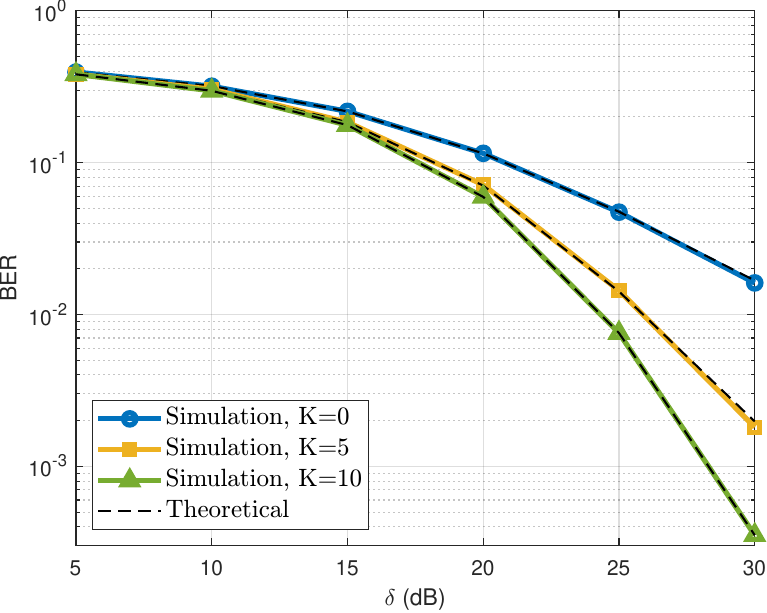}
        \text{(b)}
        \label{fig:ber_k}
    \end{minipage}
    \hfill
    \begin{minipage}{0.33\textwidth}
        \centering
        \includegraphics[width=\textwidth]{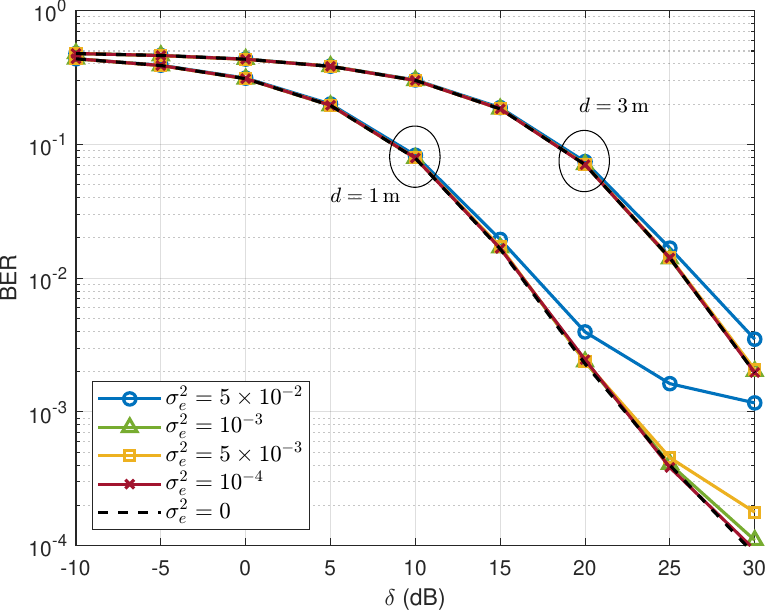}
        \text{(c)}
        \label{fig:ber_csi}
    \end{minipage}%
\caption{$\delta$ versus theoretical and simulated BER for (a) varying distances, (b) varying Rician $K$-factor values and (c) varying $\sigma_e^2$ values.} 
\label{fig:BERs}
\vspace{-1em}
\end{figure*}

\subsection{Energy Harvesting Capability}
Fig. \ref{fig:energy_harvesting} illustrates the energy harvesting capability, denoted as $z_{\mathrm{DC}}$, for diverse modulation schemes, including the TS and PS modes of the proposed JEIH-NoiseMod, 16-QAM, 16-PSK, BPSK, RG, and CSCG signals. In the PS baseline scenario, the signal power is divided with a power-splitting factor of $\rho$ for energy harvesting and $1-\rho$ for information harvesting through the power divider block before the sampling process. In this simulation, we set $N_e = 100$. The proposed JEIH-NoiseMod scheme significantly improves energy harvesting capability, particularly at shorter distances. As expected, when the distance increases due to the path loss effect, the energy harvesting capability decreases for all modulation schemes. However, JEIH-NoiseMod in TS mode consistently outperforms conventional modulation schemes. Notably, its performance is comparable to RG, which is considered ideal for energy harvesting.
\begin{table}[t!]
\centering
\caption{Simulation Parameters}
\label{tab:simulation_parameters}
\resizebox{0.65\columnwidth}{!}{
\begin{tabular}
{|p{0.25\linewidth}|c|}
\hline
\textbf{Parameter} & \textbf{Value} \\ \hline 
$R_\mathrm{ant}$ $[\Omega]$      \cite{Kim20220}         & $50$ \\ \hline
${k_2, k_4}$                   \cite{Kim20220}       & $0.0034, 0.3829$ \\ \hline
$f_c$ $[\mathrm{MHz}]$            & $433$  \\ \hline
$N$    $[\mathrm{samples}]$            & 150 \\ \hline
$d$  $[\mathrm{m}]$      & $[1, 10]$ \\ \hline
$\mu_H, \mu_L$         & $\sqrt{0.5}, -\sqrt{0.5}$ \\ \hline
$\alpha$           & $0.2, 0.3, 0.4, 0.5$ \\ \hline
$\sigma_x^2$           & $1$ \\ \hline
$\rho$ & $0.5$ \\ \hline
\end{tabular}}
\end{table}

\subsection{Error Performance}
Fig. \ref{fig:BER_Ni} illustrates the BER performance as a function of the variance ratio \(\delta\) for different values of \(N_i\). We choose $K=5$, \(d = 3\,\text{m}\) and use the \(\alpha\) values from Table \ref{tab:simulation_parameters}, where the \(N_i\) values are set to match the corresponding \(\alpha\) values.
As \(N_i\) increases, the BER improves significantly due to the more accurate averaging of noise samples. Expectedly, \(N_i = 120\) achieves the lowest BER across all \(\delta\) values, demonstrating the benefits of processing more samples for reliable detection. Compared with the conventional NoiseMod scheme and our proposed scheme JEIH-NoiseMod with $N_e=120$, there is a slight difference in terms of error performance. The reason for the slight difference is due to the number of noise samples; for the JEIH-NoiseMod cases, \(N_i\) was used, which splits the \(N\) accordingly. Expectedly, since $N$ is not split and is used only for information reception in conventional NoiseMod, a difference can be observed for various $N_i$ values that are less than $N$. Furthermore, NoiseMod uses mean values to carry information.

The BER performance for varying transmitter-receiver distances \(d\) for $K=5$ is illustrated in Fig. \ref{fig:BERs}(a) to demonstrate the effect of increasing path loss on the JEIH-NoiseMod scheme. Shorter distances result in better BER performance due to reduced path loss and stronger received signal power, where the proposed scheme showcases reasonable energy harvesting capability. The BER performance degrades as the distance increases, with \(d = 10\) meters showing the highest BER across all \(\delta\) values. Moreover, the theoretical and simulated results align closely, validating our analytical expressions.

The impact of the Rician fading parameter $K$ on the BER performance is evaluated in Fig. \ref{fig:BERs}(b), where $d = 3$ m and $N_i = 90$ samples are used. Higher values of $K$, indicating a stronger line-of-sight (LoS) component, result in improved BER performance. For \(K = 10\), the system achieves the lowest BER, demonstrating the advantage of a dominant LoS channel. When $K = 5$, the proposed system maintains satisfactory error performance. Notably, even in the absence of an LoS component ($K = 0$), the system ensures reliable communication, highlighting its robustness under severe multipath conditions.

The impact of channel estimation errors on the BER performance is illustrated in Fig. \ref{fig:BERs}(c) for distances of $d=1$\,m and $d=3$\,m under different estimation error variances $\sigma_e^2$. In computer simulations, we set $K = 5$ and $N_i = 90$ samples are used. The imperfect CSI is modeled as  
$
\hat{h} = h + \epsilon,\quad \epsilon \sim \mathcal{CN}(0,\,\sigma_e^2),
$ and $\hat{h}$ is used instead of $h$ in\eqref{eq:Mean_detector}.   
As $\sigma_e^2$ increases, only slight BER degradations are observed across all $\delta$ values, demonstrating that the proposed JEIH-NoiseMod scheme remains robust even with channel estimation errors. This resilience arises from our simple threshold-based detection rule \eqref{eq:Mean_detector}, which relies on the sample mean rather than precise channel amplitude estimates. Thus, erroneous channel coefficients have minimal impact on the decision metric. The close alignment between the perfect CSI and imperfect CSI curves confirms that the mean detector effectively mitigates the effect of estimation errors across varying distances.

Despite the inherently low bit rate characteristic of backscatter communication and IoT-based systems \cite{jameel2019simult}, our proposed scheme effectively integrates both information and energy harvesting without compromising performance. This demonstrates its ability to ensure reliable data transmission while benefiting from Gaussian signaling for energy harvesting, making it a suitable choice for low-power applications. Consequently, by utilizing the NoiseMod scheme, the proposed system maintains strong energy harvesting capability and error performance, showcasing its potential for future energy-constrained IoT applications.

\section{Conclusion}
In this paper, building on the existing NoiseMod scheme, we have introduced an innovative JEIH-NoiseMod scheme for simultaneous power and information transfer mechanism by exploiting Gaussian noise modulation for simultaneous energy and information harvesting. Embedding data in the mean of the Gaussian noise paved the way for us to propose a scheme that achieves efficient wireless power transfer while maintaining reliable communication performance, as shown through theoretical analysis and computer simulations. The results emphasize that JEIH-NoiseMod outperforms the conventional modulation schemes in terms of energy harvesting capability. In addition, it achieves promising BER performance across varying conditions, including Rician fading and path loss. Hence, it shows its potential to adapt to challenging IoT environments. Future research could focus on extending the scheme to more complex multi-user environments or increasing the data rate by exploiting other statistical properties of the Gaussian signals. Further, adaptive configurations that optimize the energy-information trade-off under diverse real-world scenarios can also be explored. One another further research can combine both energy harvesting and multiple access capabilities within a unified framework.

\bibliographystyle{IEEEtran}
\bibliography{IEEEabrv,references}


\end{document}